%% file: CF.tex
\documentclass[10pt]{amsart}
\include{russ2latex}

\date{4 May 2011}

\begin{document}
\title[Flawed Social-Network Analysis]{The Spread of Evidence-Poor Medicine\\
via Flawed Social-Network Analysis}

\dedicatory{Dedicated to the memory of David A.~Freedman}

\author{Russell Lyons} 
\thanks{Department of Mathematics, 831 E. 3rd St., Indiana University,
Bloomington, IN 47405-7106, USA.
{\tt rdlyons@indiana.edu, http://mypage.iu.edu/\string~rdlyons}}


\begin{abstract}
The chronic widespread misuse of statistics is usually inadvertent, not
intentional.
We find cautionary examples in
a series of recent papers by
Christakis and Fowler that advance statistical arguments for the 
transmission via social networks of various
personal characteristics, including obesity, smoking cessation, happiness,
and loneliness. 
Those papers 
also assert that such influence extends to three degrees of separation
in social networks.
We shall show that these conclusions do not follow from Christakis and
Fowler's statistical analyses. 
In fact, their studies even provide some evidence against the existence of
such transmission.
The errors that we expose arose, in part, because the assumptions
behind the statistical procedures used were insufficiently examined, not only
by the authors, but also by 
the reviewers.
Our examples are instructive because the practitioners are highly reputed,
their results have received enormous popular attention,
and the journals that published their studies are among the most respected
in the world. An educational bonus emerges from the
difficulty we report in getting our critique published.
We discuss the relevance of this episode to understanding statistical
literacy and the role of scientific review, as well as to reforming
statistics education.
\end{abstract}

\maketitle 
\renewcommand{\thefootnote}{\arabic{footnote}}

\def\fplp{FP$ \to $LP}
\def\lpfp{LP$ \to $FP}
\def\mut{FP$ \leftrightarrow $LP}
\def\P{{\bf P}}
\def\E{{\bf E}}


\bsection{Introduction}

For at least 130 years, it has been
common knowledge that statistics are widely abused.
Less well known among the public is that professional publications even in
top medical journals routinely, though unwittingly, misuse statistics.
The corollary that top journals do not serve as rigorous judges of quality,
due to lack of statistical competence, is not often discussed.

We illustrate the latter two themes
in this paper by presenting some cautionary examples of somewhat
sophisticated recent statistical analyses that were flawed by
insufficient attention to assumptions and misinterpretation of results.
Novel techniques were used to analyze social
networks. The results of these analyses were
published in the most respected medical journals and have become rather
famous, even outside academia. However, both elementary statistical errors
and more advanced errors undermine these analyses to such an
extent that little can be deduced from the original studies---except
that we need to improve our statistics education. Despite medicine's recent
emphasis on improving the nature of their evidence, the medical field still
has a long road ahead.  

We hope that our analysis will be useful to educators, to
practitioners, and to all who have an interest in the quality of scientific
research that relies on statistics.
With such audiences in mind, we have endeavored to explain
our analysis as carefully as possible, while minimizing mathematical
derivations. 

The statistics in question come from
a series of recent papers \cite{CF:obes,CF:smoke,CF:happy,CFC:alone} by
Christakis and Fowler (C\&F), who analyzed network data coming from the
Framingham Heart Study. This long-running observational study collects not
only physical health information, but also other personal characteristics,
including elements of the social network of participants. 
C\&F analyzed new data via new statistical techniques, 
leading to two major inferences:
\begin{enumerate}
\item There is a process of infection or contagion within this
social network that transmits various personal characteristics, including
obesity, smoking cessation, happiness, and loneliness. 
\item Such transmission occurs up to three steps in the network, providing
evidence of a universal `` `three degrees of
influence' rule of social network contagion" \cite{CFC:alone}.
\end{enumerate}

C\&F's studies have received considerable acclaim 
in the popular press and in society at large. For 
example, their study on obesity was reported on the front page of {\it The
New York Times}, above the fold, and was at some time
e-mailed from the website more than any other article but one that day.
Both authors were named one of the ``Top 100 Global Thinkers" in 2010 by {\it
Foreign Policy} magazine.
Rudolph Leibel, a member of the Institute of Medicine of the National
Academy of Sciences, said of C\&F's paper \cite{CF:obes} on obesity that
``It is an extraordinarily subtle and sophisticated way of getting a handle
on aspects of the environment that are not normally
considered" \cite{Kolata}.
Daniel Kahneman, a Nobel-prize winner, said of C\&F's paper \cite{CF:happy}
on happiness that
``It's extremely important and interesting work "\cite{Belluck}.
Considerable professional success has attended their work, with large
grants coming their way; the largest to date is for \$11,000,000 from the
National Institute of Aging.
Their conclusions have also been disseminated
via a popular book \cite{CF:conn}, which has been translated into twenty
languages. 

Despite such accolades, we shall establish that
both of their major claims are unfounded.
That is, while the world may indeed work as C\&F say, their studies do not
provide evidence to support such claims.
Moreover, parts of their studies even suggest that their claims of
transmission are untrue.

In the remainder of this introduction, we present a summary of their
evidence and a summary of our arguments against it.
Later sections provide details.

All of C\&F's papers in this series use similar methods, so
for brevity, we refer only to their obesity study.
The Framingham Heart Study has about
12,000 participants, who are examined every few years.
About 5,000 of the participants are in the ``Offspring Cohort".
It is those in the Offspring Cohort whose obesity is analyzed in relation
to the obesity of all 12,000 participants.

C\&F start by finding statistical associations between the obesity of
friends in the Framingham network: To oversimplify, a person's
friends are more likely to be obese if the person himself is obese.
The associations that C\&F analyze are calculated from
statistical models whose parameters are estimated by using
the observational data. This source of C\&F's associations
is crucial to their argument and decisive for our critique.

C\&F argue that these associations are not mere associations, but
measure causal effects.
The two primary reasons the associations might not be causal are 
{\em homophily} (or selection), which is the fact that people
tend to associate with others like themselves, and a {\em shared
environment} (also called ``confounding" or ``contextual influences" by
other researchers). C\&F call the causal effects
{\em induction} (also called ``influence" or ``endogenous social effect" by
others) that they liken to
a transmission process.
C\&F deduce induction indirectly 
by ruling out the possibilities of homophily and
shared environment; they provide speculation, but not evidence,
for how such induction might work.

For concreteness in our explanations, suppose that Frank is a study
participant in the Offspring Cohort.
C\&F use a logistic regression to model
the probability of Frank's obesity at a given exam.
The important variables used in the model are
Frank's obesity status at the previous exam and the obesity status---both now
and at the previous exam---of those (such as, say, Linda) to whom Frank is
connected in his social network. 

C\&F argue against the homophily explanation because 
their logistic
regression model included a term for Linda's obesity status at the previous
exam. Since the model produces the associations to be studied,
they are supposed to be net of any effects of homophily.

C\&F argue against the shared environment explanation as follows.  
Consider two friends, Frank and Linda. C\&F have ``directional" information
on friendships:
Each participant was asked to name one close friend.
Suppose Frank named Linda as his ``closest" friend, but not vice versa.
C\&F find that if Linda becomes obese, then Frank's chance of becoming
obese himself increases by 57\% relative to what it would be if Linda did
not become obese.
On the other hand, suppose that
Linda named Frank as {\em her} closest friend, yet
Frank did {\em not} name Linda as {\em his} closest friend. In this
case, if Linda becomes
obese, then Frank has only a 13\% increased chance of becoming obese. 
Since 57\% is far different from 13\%, C\&F
contend that this asymmetry rules out a shared environment between Frank and
Linda as a cause of their associated obesity.
C\&F conclude that having accounted for or ruled out the other
possible explanations for the observed associations in obesity, it must be
induction that produces these associations.

In order to establish their
three-degrees-of-influence rule, C\&F compare the network data they have
to random networks, where they change who is obese, while maintaining the
existing social ties. By comparing statistical 
associations in the actual network to
those in the random networks, they find that obesity is significantly
associated out to three degrees and not further.

While the influence of friends' obesity on others depends on social
distance in this way, according to C\&F, it does not depend on geographic
distance, even when the friends involved rarely see each other.
C\&F also say that obesity spreads to a friend of a friend (or even to a
friend of a friend of a friend)
without the intermediate friend(s) becoming obese \cite{NYTMag}.

However, the arguments
given to substantiate C\&F's claims are not sound, primarily because of
two kinds of errors: 
\begin{enumerate}
\item C\&F use
statistical models that contradict their data, as well as their
conclusions.
\item 
Even if one accepts C\&F's statistical models and
tests, C\&F interpret the results incorrectly.
\end{enumerate}

As we have noted,
the increases in obesity risk
reported above do not arise from calculations based directly on the data.
Rather, they arise indirectly from the data:
They result from statistical models that were fitted to the
observational data.  By the nature of a statistical model, the numbers
above, 57\% and 13\%, come with uncertainties.  C\&F say that these numbers
are statistically distinguishable. However, when we look more closely (in
critique (1) of \rref s.direct/), we shall
see that they are in fact {\em not} distinguishable---due to the large
uncertainties inherent in them.  
We shall also demonstrate (in critique (2) of \rref s.direct/) that their
addition of a lagged obesity term in their models does not
properly control for homophily; rather than subtract the effect of
homophily, if anything, it amplifies it. 
Moreover, a closer examination of the idea of directional associations
will show (in critique (3) of \rref s.direct/)
that the proposed differences are actually
consistent with all three types of explanation: homophily, shared
environment, and induction.  In sum,
C\&F have not shown that they can distinguish
among the three possible explanations.

We shall examine the first category of error in \rref s.model/; as it is
the most technical aspect of our analysis, we reserve it for last.
C\&F's statistical models will turn out to
have serious problems due precisely to the network effects C\&F hope to
analyze.
For example, the asymmetry discussed above, produced by their model and
intended to rule out a shared-environment explanation, turns out to be
mathematically inconsistent with their model.
How can the model produce a result that is inconsistent with itself? It is
because 
C\&F's method of estimation of their model is inapplicable to their model.
All these problems cast doubt on C\&F's reported numbers.
Moreover, as noted above, C\&F provide other evidence that
associations persist in the face of geographic separation; this suggests
that homophily is, in fact, playing the major role.
In our view, the most important task of C\&F is to show that homophily does
not explain their associations. 
For a simple example showing how homophily relates to
shared changes in health, suppose that Sally gets cancer. Then her friends
are more likely to have gotten cancer than those who are not Sally's
friends. Why?
Because Sally is likely old and so are her friends.
(Of course, in this example, one can control for age. The difficulty in
general is to control properly for all confounding factors, including the
unknown ones.)

It is true that the three-degree
rule exists in the network data that C\&F use. However, this is partly 
due to the nature of their data, which
is sparse.
For example, in many cases, friends
of friends will be friends, but this is not recorded in their data. The
network assembled from this data, therefore, is likely to mislead.


Following our critique of C\&F's work, we consider the implications
for quality control at top journals in \rref s.review/. We also describe
briefly the difficulties we had in getting our critique published and the
attitudes we encountered from top journals towards critiques. In our last
section, we place this episode in a general context of a misplaced faith in
statistical models, illustrated with quotes from distinguished critics.
We urge that statistics education place much more emphasis on critical
thinking.

\bsection{Directionality}
\label{s.direct}

%
We begin with a critique of C\&F's argument against the shared environment
explanation. Their argument is based
on perceived directional differences in friendships. To
understand the issues, we must review a key trait of their studies. 

As we have said,
certain participants, the Offspring Cohort, are chosen to be the focus
of analysis; they are called ``focal participants" (abbreviated FP) in
\rref b.CFC:alone/, and are called ``egos" in the other studies. The
participants to whom they are linked by a tie of friendship, family,
workplace, or neighborhood are called ``linked participants" (abbreviated
LP) in \rref b.CFC:alone/, and called ``alters" in the other studies. Some
LPs are also FPs. Thus, FP is an absolute term, while LP is relative to the
FP. 
As we said above, 
each participant was asked to list one close friend.
(Some people listed more than one, despite the instructions.)
The friendship data in the Framingham Heart Study consists of the record of
those answers.
This leads to the key property that friendship
ties are directional, from FP to LP or from LP to FP.
In case each names the other, then the tie goes both ways.
Some ties are between two LPs, neither of whom is an FP, so those
ties are not included in most of C\&F's analyses.
Furthermore, only ties to people who also were in the Framingham Heart
Study were included in C\&F's analyses.
In the case of friends, for example, those included
amounted generally to less than 1/4 of all named friends: see
\cite[supplement, Table S2]{CF:smoke}.

Here is how C\&F explain the directional differences in \rref b.CF:obes/:

\technical

If an ego stated that an alter was his or her friend, the ego's chances of
becoming obese appeared to increase by 57\% (95\% confidence interval
[CI], 6 to 123) if the alter became obese.  However, the type of friendship
appeared to be important. Between mutual friends, the ego's risk of obesity
increased by 171\% (95\% CI, 59 to 326) if an alter became obese. In
contrast, there was no statistically meaningful relationship when the
friendship was perceived by the alter but not the ego (P = 0.70). Thus,
influence in friendship ties appeared to be directional.

\dots

\noindent
the findings regarding the directional nature of the effects of friendships
are especially important with regard to the interpersonal induction of
obesity because they suggest that friends do not simultaneously become
obese as a result of contemporaneous exposures to unobserved factors.  If
the friends did become obese at the same time, any such factors should have
an equally strong influence regardless of the directionality of friendship.
This observation also points to the specifically social nature of these
associations, since the asymmetry in the process may arise from the fact
that the person who identifies another person as a friend esteems the other
person.

\endtechnical

In order to discuss this argument, it will be useful to abbreviate a
friendship tie as
\fplp\ when the FP named the LP but the LP did not name the FP;
\lpfp\ when the LP named the
FP but the FP did not name the LP; and \mut\ when the naming was
mutual. 
Thus, C\&F are saying that causality is the best explanation for the
differences among 171\% for \mut, 57\% for \fplp, and 13\% for
\lpfp. 

We claim that C\&F's argument from
directional differences has
the following three problems,\footnote
{Versions of these three problems were mentioned briefly in the editorial
\cite{StepRoux}. The latter two were also discussed in the letter
\cite{Morgan}. A theoretical discussion related to the second point, whether
it is even possible to control for homophily without making assumptions,
is given by \rref b.ShaTho/.}
which we discuss in turn:
\begin{enumerate}
\item The differences are not statistically significant.
\item C\&F's argument that the differences are net of homophily is
incorrect.
\item The differences are consistent with all three possible explanations.
\end{enumerate}

(1) The first problem is that the differences are not statistically
significant.  Let us consider carefully their reasoning: C\&F estimate an
\fplp\ increased obesity risk of 57\% and an \lpfp\ increased obesity risk
of 13\%. However, they accept that their estimates are not precise. They
feel 95\% confident that the former lies in the interval from 6\% to 123\%,
while the latter, being statistically insignificant,
might well be 0\%. 
Since 0\% does not lie in the interval $[6\%, 123\%]$,
they infer that the two risks are different. 
But this reasoning exemplifies a statistical error that is common in many
studies and that occurs throughout C\&F's (\rref s.further/ of the Appendix).
The error is to mistake a number for 0 when one has learned only that the
available evidence is too imprecise to distinguish the number
from 0. 
In the present case, the estimate 13\% for the \lpfp\
risk has a CI that seems to be $[-28\%, 68\%]$.  C\&F take the ``true
value" to be 0\%, but there is no
reason to take the ``true value" to be 0\%.  Their estimate is 13\% and
13\% itself falls {\em in} the CI for the \fplp\ risk. To compound the
error, 57\% also falls in the CI for the \lpfp\ risk.
This means that C\&F's numbers do not
distinguish the associations in the two directions.  The observed
differences could be due to chance, according to C\&F's
technique.

This same error regarding statistical significance for directional
estimates
occurs in each of C\&F's papers; it is summarized in
\rref f.CIslabs/. (See Table 1 of \rref a.dtable/ for the
numerical estimates and intervals.)

\begin{figure}[htb]
\begin{center}
\includegraphics{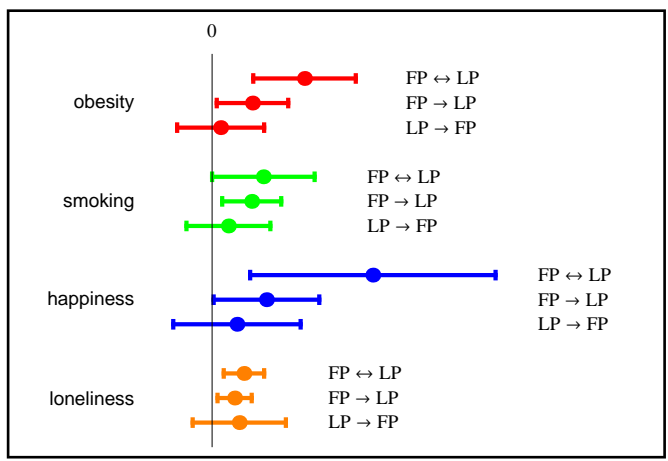}
\captionsetup{width=5.3in}
\caption{Coefficient 
estimates and 2 SE (95\%) confidence intervals for directional
effects. For each study, the order from top to bottom is (1)
mutual friendship,
(2) FP named LP, then (3) LP named FP. The CIs overlap so much that one
cannot infer that the
differences are statistically significant. Sources:
\cite[suppl.~p.~3]{CF:obes};
\cite[suppl.~p.~18]{CF:smoke};
\cite[suppl.~p.~9]{CF:happy};
\cite[pp.~983--984]{CFC:alone}.
}
\label{f.CIslabs}
\medskip
\end{center}
\end{figure}

A technical note:
C\&F are comparing coefficients from different models. Therefore, they
cannot estimate the difference between these coefficients. They would be
able to make an inference on the difference of two coefficients
if they had a valid
model that contained both coefficients. 
We don't know such a model and, for the general reasons discussed in
Section 7, we are skeptical that one exists. Putting such skepticism aside,
if we wished to construct such a
model, we would need access to the data. The social network data for the
Framingham Heart Study was assembled by C\&F from 
hand-written data, but C\&F have not
made this available to others. This also prevents the most basic type of
replication \cite{replicate} and can keep errors hidden \cite{BagCoo}.
In any case, given what C\&F have, they do not have reason to
infer that these differences are statistically significant.

(2) Suppose we ignore this inferential difficulty
and allow C\&F their directional
differences; after all, there is indeed a clear pattern in the estimates.
According to C\&F, these differences rule out confounding.
What about homophily?
C\&F counter this explanation as follows.
The numbers above (such as 57\% and 13\%) arise from logistic regression
models.
C\&F \cite{CF:obes} say,
``Our models account for homophily by
including a time-lagged measurement of the alter's obesity."
That is, in the equations predicting the FP's current obesity, there is a
variable that indicates whether the LPs were obese in the previous exam.
Therefore, the risks that C\&F analyze are supposed to be net of
whatever effects may be due to homophily.
C\&F do not give a separate argument against homophily. 
Their reasoning hinges, then, on whether the lagged term properly controls
for homophily. 
Let us look.

Their model has two related terms, one for the LP's current
obesity and the other for the LP's lagged obesity, i.e., the LP's obesity
status at the previous exam. The current obesity is used to measure
``effect" on the FP's obesity, while the lagged obesity is used to
``control" for homophily. One might argue that the reverse (if either)
should be used,
as causal effects require a time difference. However, either choice leads
to disquiet when we use the estimates C\&F give for the two corresponding
coefficients, as these two coefficients sum to approximately 0 \cite
[suppl., Tables S1, S2, S3]{CF:obes}. In particular, they have opposite
signs. Thus, if the
lagged obesity were used for ``effect", we would conclude that the social
network {\it inhibits} the spread of obesity, while if the lagged obesity
were used for ``control", as C\&F do, then we would be left with the puzzle
that homophily affects the FP and the LP in opposite ways.
Should we find such opposite effects too unsettling, then to the extent
that the lagged term relates to homophily, we would conclude
that rather than taking away the effect of homophily, the term
has amplified its effect.

We remark that Cohen-Cole and Fletcher
\cite{CCF:contag,CCF:detect} also felt
that C\&F had not controlled properly for homophily. 
They attempted to show the unreliability of C\&F's work by deducing implausible
conclusions from similar modeling techniques and by showing how the conclusions
change with different controls. 
C\&F \cite[full version at the authors' websites]{CF:peer} responded by noting
that their critics found only statistically insignificant results and by a
simulation.
Another difficulty in C\&F's work was
pointed out by \cite{NoelNyhan}:
When
friendships change in ways related to homophily, then
estimates of the effects of variables other than homophily can be biased.

(3) The third problem is that directional differences are actually
consistent with all three considered explanations, i.e., induction,
homophily, and environment.
Consider the three types of ties, \mut, \fplp, and \lpfp, and, for each
type, the
possible correlations of the LP's obesity with the obesity of the FP.
As shown in \rref f.CIslabs/,
C\&F find that the strength of these three correlations are different and
in order of most to
least. They say that this eliminates the possibility
that these correlations are due to a
shared environment.
We say that one expects this same ordering of strength of correlation
whether the correlations are due to induction, homophily, or even shared
environment.
Furthermore, this is true regardless of whether one finds these
correlations due to modeling or other reasons: this is a general phenomenon
arising from the choice of whom to correlate with the FP.
C\&F have argued the case for induction (who ``esteems" whom); we explain
why the same holds for homophily and shared environment.

Consider the following hypothetical situation.
Imagine that each individual names as a friend the other person whose
characteristics (covariates) are objectively 
closest to his own. See \rref f.ties/ for a representation, to be explained
further below.
If we are thinking about homophily, then this naming process represents people
selecting each other based on similar characteristics. If we are thinking
about shared environment, on the other hand, then this process represents
people making friends with their closest neighbors.

One of the individuals is Frank, an FP, who names Linda, an LP, as his friend.
How correlated is Frank's obesity with Linda's as a function of their type
of friendship?
By definition, Linda is closer to Frank than anyone else, including
those who named Frank as their closest friend.
If distance represents degree of homophily, then this means Linda is more like Frank
than anyone else, while if distance arises from location, then Linda shares
more of Frank's environment than anyone else.
In each of these two cases,
Linda is more correlated with Frank 
than are those others who name Frank (the \lpfp\ ties).
Finally, if it happens that Linda also named Frank reciprocally (so
their tie is of \mut\ type), then
this pair of individuals is especially close to each other and thus Linda
is 
even more correlated with Frank.
Thus, we see in this hypothetical situation precisely the kind of directional
differences C\&F find: \mut\ ties have the strongest associations, followed
by \fplp\ and then \lpfp. 
In sum, the directional differences C\&F found are just what
one would expect to see for all three types of explanations; the
differences do not distinguish among the explanations. 

\begin{figure}[hbt]
\begin{center}
\includegraphics[height=4truein]{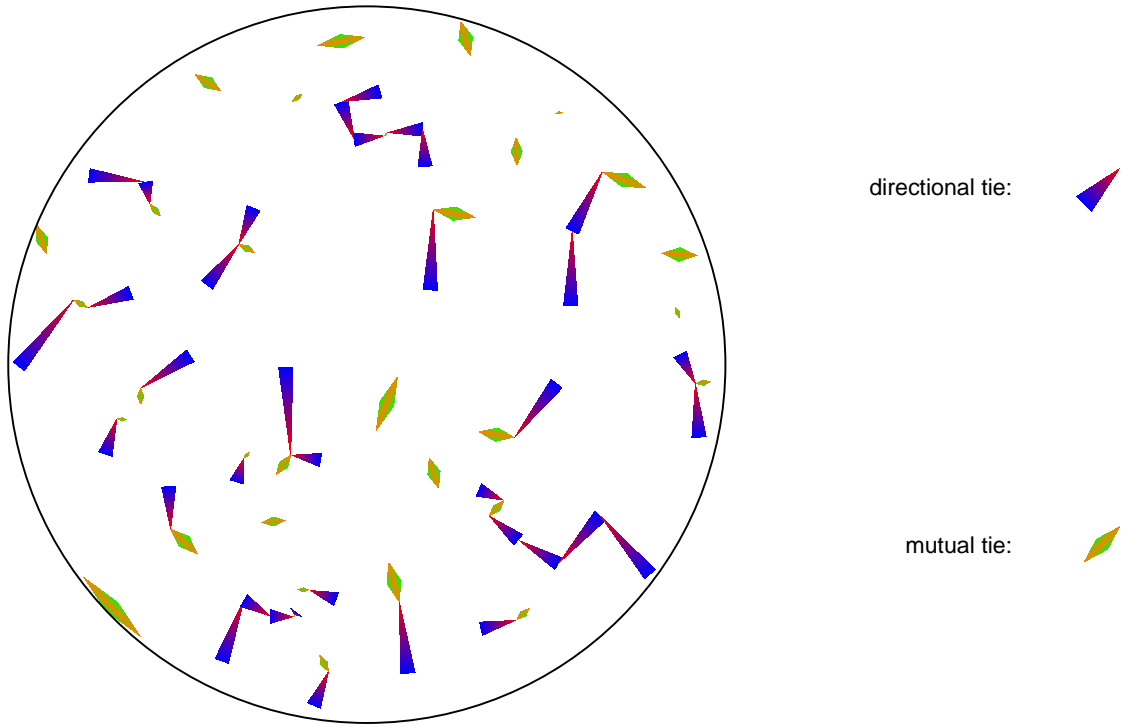}
\caption{100 random locations in a disc, each pointing to its nearest
neighbor. Locations that point to each other are usually especially close
to each other.}
\label{f.ties}
\medskip
\end{center}
\end{figure}

We now discuss \rref f.ties/ in order to elucidate
why
mutual friends are especially close to each other. Consider the
following chance model for the above hypothetical situation.
Let $B$ be a ball in a high-dimensional
space. The location of an individual in $B$ represents various of his
covariates.
Imagine that individuals are independently uniformly distributed in
$B$. Now each person
names one other as a friend, namely, that person who is closest to him. 
\rref f.ties/ shows this in two dimensions.
One can easily prove
mathematically that the distance between mutual friends
is stochastically smaller
than the distance between non-mutual friends.
(This means that for every number $d$, the probability that the distance
between mutual friends is less than $d$ is at least the probability that
the distance between non-mutual friends is less than $d$.)
Thus, mutual friends are generally closer to each other than are non-mutual
friends, as is apparent visually in the figure.




\bsection{Random Networks}
\label{s.rn}

We now consider the methods and meanings of C\&F's statistical calculations.
They use two methods across their papers: one consists of varying values in
the given network, while the second consists in making 
regressions. 
The first method leads to C\&F's 
three-degrees-of-influence rule, while the second method leads to the
estimates and CIs discussed in the preceding section. Although it
would be logical to discuss the regressions now, they are much more
technical, so we defer that discussion until after we discuss the random
networks in this section.

In \rref b.CF:obes/, the authors preserve the network, but randomly
redistribute the incidence of obesity (preserving the same number of obese
individuals). 
By comparing the actual network to the randomly generated networks, C\&F
demonstrate that statistical associations of obesity between pairs of
people extend to three degrees of separation in the observed network.
The same result holds for smoking cessation, happiness, and loneliness.
This
is a reasonable method to summarize the structure that exists in the observed
network as it relates to the characteristic of interest.

However, as we noted already, the data used by C\&F is incomplete
and thus the network treats some
people as not friends when in reality they are friends.
Only 45\% of the 5124 FPs named a friend in
the Study. There were 3604 
unique observed friendships in total \cite
{CF:obes,CF:happy}, but not all were among those named at any one
time.
This means that the average current number of
friends reported was about 0.7 per FP.
Thus, the network data concerning friends,
in particular, is quite thin.
Therefore, the three-degree pattern, while present in the network assembled from
the data used, has not been
demonstrated for the real world.

This random-network analysis is unrelated to the cause of the statistical
associations; C\&F turn to regression models to argue their causal
conclusions. But since C\&F find in \cite{CF:obes,CF:smoke} that the
associations in their networks are essentially unrelated to geographic
distance, they have in fact given evidence that the associations of obesity
and smoking are due to homophily, more than to a shared environment, and
unlikely due to induction.

A further problem is that
the language C\&F use blurs the distinction between the model and reality.
For example, \cite{CF:obes} 
reports that ``the risk of
obesity among alters who were connected to an obese ego (at one degree of
separation) was about 45\% higher in the observed network than in a random
network."\footnote 
{The caption of their
Fig.~3 differs from the text as to what they measured.
The caption interchanges ``alter" and ``ego" and
reports this as ``Relative Increase in Probability of Obesity in an
Ego if Alter becomes Obese". 
The same conflict of description occurs with Fig.~2 in \rref
b.CF:smoke/.}
In \rref b.CFC:alone/, C\&F write that ``a person's loneliness depends not
just on his friend's loneliness but also extends to his friend's friend and
his friend's friend's friend. The full network shows that participants are
52\% (95\% CI $=$ 40\% to 65\%) more likely to be lonely if a person to
whom they are directly connected (at one degree of separation) is lonely."
This sounds very much like predicting what would happen in the real world
if a friend became lonely, which, after all, is a main goal of the paper.
Indeed, C\&F frame this particular figure
in terms of ``the `three degrees of
influence' rule of social network contagion that has been exhibited
for obesity, smoking, and happiness" \cite{CFC:alone}.
Further reinforcing the idea that C\&F are making a prediction, 
the CI appears to quantify the uncertainty in the prediction. 
However, this 52\% is not a risk of friendship, nor is it a prediction
about the real world, nor did it involve comparisons across time.
It is merely a numerical comparison of the observed
network to a certain random network. 
Likewise, the CI is not actually a confidence interval:
Confidence intervals aim to contain the true unknown parameter and are obtained
by random sampling from the true population, whereas here,
we know the real network and randomly sample the imaginary network, which,
by design, is not realistic.
This blurring of model and reality can mislead readers. Accordingly,
the editorial
\cite{Sainsbury} commented on \cite{CF:happy}
that ``the size of
the influence of distant friends (friends of friends' friends; 5.6\%) seems
overly large when the influence of a happy friend is only 14\%."
The figures quoted here by \cite{Sainsbury} come directly from C\&F's
random networks and so do not represent influences.

\bsection{Modeling}
\label{s.model}

The bulk of the numbers produced by C\&F arise from an abundance of
logistic or linear regression models, intended to describe and
explain the observed associations. 
C\&F have not done an experiment,
nor run a so-called natural experiment, and
they do not have enough data for multi-dimensional
cross-tabulation; this is why
they turn to statistical models \rref b.SM:TaP/.

Use of a statistical model requires C\&F to make
assumptions about what
the data would look like if either they had an experiment or they had much
more data. If these assumptions are wrong, then C\&F's conclusions may be
invalid or misleading. C\&F pay very little attention to their
assumptions, but they are crucial for the validity of their methods. We
shall examine only a few of those assumptions here.
Notably, we shall find that their regression models contradict
their data and their conclusions about directional effects.

In order to see clearly what C\&F assume, it is important to describe
precisely their models. The first time that C\&F stated their
models was in \cite{CF:aoas}, where they were invited to discuss the
statistical foundations for their work.
C\&F aim to reveal causation by technical means
(they do not claim to have observed an induction mechanism),
and only a technical examination can reveal fully the flaws.

Let $Y_{i, t}$ be the indicator that individual $i$ is obese at time $t$.
(An indicator is 1 if the event is true, and 0 otherwise.)
These times can be any integer from 1 to 7; they correspond to exam periods
(called ``waves"), which occurred every few years.
Let $W_n(t)$ be the indicator that $t = n$; let $A_{i, t}$ be the age of
$i$ in years at time $t$; let $F_i$ be the indicator that $i$ is female;
and let $E_{i, t}$ be the number of years of education of $i$ at time $t$.
Abbreviate by $C_{i, t}$ the collection of indicator variables
$Y_{j, s}$ for all pairs $(j, s) \ne (i, t)$.
Different models arise by considering various sets 
$T_t$ of ties at times $t$, as well as by changing the covariates
listed above.
For given sets $T_2$, $T_3$, $T_4$, $T_5$, $T_6$, and $T_7$
($T_t$ might equal, say, all mutual ties among FPs
that existed at both times $t$ and
$t-1$), C\&F posit a
system of simultaneous equations for the joint distribution of $Y_{i, t}$: 
there are some numbers $\alpha$, $\beta_k$, $\gamma_n$, $\delta_k$ such that
for all $2 \le t \le 7$ and all $(i, j) \in T_t$, we have\footnote
{One might also make the probabilities on the left-hand side of the
equation conditional on the other covariates, but we shall treat them as
non-random for brevity.}
$$
\log \frac
{\P[Y_{i, t} = 1 \mid C_{i, t}]}
{\P[Y_{i, t} = 0 \mid C_{i, t}]}
=
\alpha + 
\Big(\beta_1 Y_{j, t} + \beta_2 Y_{j, t-1}\Big) +
\beta_3 Y_{i, t-1} + \sum_{n=3}^7 \gamma_n W_n(t) + \delta_1 A_{i, t} +
\delta_2 F_i + \delta_3 E_{i, t}
\,.
$$
The two terms in parentheses on the right-hand side are the key terms.
C\&F's main interest is in estimating $\beta_1$, which they call the
``effect" of the LP's current obesity on the FP's current obesity.
The summand $\beta_2 Y_{j, t-1}$ is supposed to control for homophily.
Change in obesity is represented by having the FP's obesity status at the
previous exam, $Y_{i, t-1}$, on the right-hand side of the equation.
The rest of the covariates on the right-hand side are supposed to control
for time and personal characteristics.

For example, for mutual friends, C\&F estimate $\beta_1 = 1.19$ with an SE
of 0.33, which they translate to an increased risk of 171\% with a 95\% CI
of $[59\%, 326\%]$.

A fundamental problem is that there are too many equations: one for each
tie. That's usually more than the number of $Y_{i, t}$.
If $i$ names more than one $j$ at time $t$, then 
$\Big(\beta_1 Y_{j, t} + \beta_2 Y_{j, t-1}\Big)$
must be the same for all such $j$ because that's the only thing that
changes in the equation when $j$ changes.
Thus, the model contradicts the data---unless $\beta_1 = \beta_2 =
0$, in which case no individual affects any other at any time.


In fact, it turns out that 
$\beta_1 = 0$ regardless of the data.\footnote{In a different context, this
same conclusion, that $\beta_1 = 0$, is proved in \cite{Heckman}; see also
\cite{Tamer}.} That is, C\&F's model contradicts their conclusions
concerning directionality.
To see this, let the set $T_t$ of ties consist of \fplp\ ties (as C\&F
do for estimating the ``risk" of \fplp\ ties).
Suppose $(i, k), (k, m) \in T_t$, where $m \ne i$.
Let $D_{i, k, t}$ be the collection of indicator variables $Y_{j, s}$ for
all pairs $(j, s) \ne (i, t), (k, t)$.
We may calculate $\log \big(\P[Y_{i, t} = 1, Y_{k, t} = 1 \mid D_{i, k, t}]
/\P[Y_{i, t} = 0, Y_{k, t} = 0 \mid D_{i, k, t}]\big)$ in two different
ways; equating them yields 
\begin{align*}
\log \frac
{\P[Y_{i, t} = 1 \mid Y_{k, t} = 1, D_{i, k, t}]}
{\P[Y_{i, t} = 0 \mid Y_{k, t} = 1, D_{i, k, t}]}
&+
\log \frac
{\P[Y_{k, t} = 1 \mid Y_{i, t} = 0, D_{i, k, t}]}
{\P[Y_{k, t} = 0 \mid Y_{i, t} = 0, D_{i, k, t}]}\\
&=
\log \frac
{\P[Y_{k, t} = 1 \mid Y_{i, t} = 1, D_{i, k, t}]}
{\P[Y_{k, t} = 0 \mid Y_{i, t} = 1, D_{i, k, t}]}
+
\log \frac
{\P[Y_{i, t} = 1 \mid Y_{k, t} = 0, D_{i, k, t}]}
{\P[Y_{i, t} = 0 \mid Y_{k, t} = 0, D_{i, k, t}]}
\,.
\end{align*}
Use of the model equation to evaluate each of these four logarithms (use
the equation for $(i, k)$ when predicting $Y_{i, t}$ and the equation for
$(k, m)$ when predicting $Y_{k, t}$)
yields an
equation in which all terms cancel but one---leaving $\beta_1 = 0$.

Since \cite{CF:smoke} and \cite{CF:happy} use these same kinds of models,
those papers share these same problems.
The paper \cite{CFC:alone} uses mostly 
linear regressions rather than logistic regressions.
These regression equations are almost the same, but now the
response variable, $Y_{i, t}$,
is the number of days per week
$i$ is lonely at time $t$. In \cite{CFC:alone},
C\&F posit that each tie $(i, j)$ at time $t \in
\{6, 7\}$
satisfies
$$
Y_{i, t} 
=
\alpha + 
\Big(\beta_1 Y_{j, t} + \beta_2 Y_{j, t-1}\Big) +
\beta_3 Y_{i, t-1}
+ \gamma_7 W_7(t) + \delta_1 A_{i, t} +
\delta_2 F_i + \delta_3 E_{i, t}
+ \varepsilon_{i, j, t}
\,,
$$
where $\varepsilon_{i, j, t}$ is an error term that (presumably) is independent
of the other variables on the right-hand side, as well as of all variables
at time $t-1$, and has mean 0.

This model also contradicts the data unless $\beta_1 = \beta_2 = 0$.
To see this, take the expectation
conditional on time $t-1$ and thereby eliminate the error terms.
If 
$
(i, j), (j, i), (j, k), (k, j) \in T_t
,
$
where $k \ne i$,
then we get 4 equations for
the 3 unknowns, these 3 unknowns having the form
$
\E[Y_{m, t} \mid \hbox{time } t-1]
$
for $m = i,j,k$.
Since the 4 equations have a solution, we get an equation that the data at
time $t-1$ must satisfy. Some algebra shows that
to prevent a contradiction, this equation implies that
$\beta_1 = \beta_2 = 0$.

\bsection{Model Estimation}

C\&F estimate their 12 coefficients via a method known
as generalized estimating equations (GEE). This method is designed for
repeated measures or other sorts of dependencies, but itself comes with
some assumptions \cite[Theorem 1]{LZ}. One assumption is independence among
subjects or among groups of subjects. Since C\&F have not clearly
delineated their use of GEE, we must guess how they are using it
from their descriptions such as ``We
used generalized estimating equations to account for multiple observations
of the same ego across examinations and across ego-alter pairs. We assumed
an independent working correlation structure for the clusters" 
\cite{CF:obes} and ``Models were estimated using a general estimating
equation with clustering on the focal participant and an independent
working covariance structure" \cite{CFC:alone}. This seems to mean that
all the measurements on each FP were a single cluster, or group. If so,
however, then these groups are not independent; indeed, C\&F
wish to make conclusions about their dependencies. The ``working
covariance" seems to relate to the different measurements in
time. 
Yet GEE relies on having a large number of independent groups.

Moreover, there is a peculiar twist to the equations of C\&F's models due to
the interest in the social network:
the dependent variables (the $Y_{i, t}$) appear not only on the
left-hand sides of the equations, but also on the right-hand sides. This is
not part of the literature regarding GEE, at least to our knowledge.
It is possible that C\&F's estimation method could work when the model
holds, but we would need to see mathematical proof or relevant
simulation results.

In any case, we do have enough information to know
that GEE does not work for C\&F's model: Since the
model implies that $\beta_1 = 0$, yet C\&F
estimate that $\beta_1 \ne 0$, it follows that their estimation method is
faulty.


\bsection{The Role of Review}
\label{s.review}

Both of C\&F's first two papers were published in the world's top medical
journal, the {\it New Engl.\ J.\ Med.} Their third paper was published in
{\it BMJ}, another very highly respected medical journal. Their fourth
paper was published in the {\it J. Pers.\ Soc.\ Psychol.}, a top journal
and the flagship journal of the American Psychological Association.
After we had completed our analysis of those four papers, two more based on
the same data appeared:
the fifth \cite{CF:alcohol} in {\it Ann.\ Intern.\ Med.}, again a very
highly respected journal, and the sixth \cite{CF:depression} in {\it Mol.\
Psychiatry}, a top journal in psychiatry.
We leave as an exercise to the reader to spot in these last two papers the
same errors we have recounted here.

Given the fundamental errors we have described, what can we conclude about
the process of peer review at these top journals? 
Altman \cite{Alt:rev}, currently
the senior statistics editor at {\it BMJ},
gave a personal account as a statistical
reviewer of submissions to
medical journals, as well as a table summarizing some studies on the
quality of statistics in published medical articles. His bleak assessment:
``The main reason for the plethora of statistical errors is that the
majority of statistical analyses are performed by people with an inadequate
understanding of statistical methods. They are then peer reviewed by people
who are generally no more knowledgeable. Sadly, much research may benefit
researchers rather more than patients, especially when is carried out
primarily as a ridiculous career necessity."

Problems with peer review have long been known and several remedies
have been proposed. One remedy has even been shown to fail: see \cite{FTCFL}.
We propose a new solution below, based partly on our experiences in 
getting the present critique published.
%
One can find several anecdotal reports on the web about the policies of top
scientific journals regarding critiques, but we are not aware of any study
of the issue. 
Our experiences matched the anecdotes we saw
and seem informative. 

We first submitted our critique to the {\it New Engl.\ J.\ Med.},
but it was rejected without peer review. The journal declined to give
a reason when asked.
We next submitted to {\it BMJ}, but it was again rejected without peer
review. This journal did, however,
volunteer that ``We decided your paper was probably better
placed in a more specialist journal." It is interesting to note
that the same issue of {\it BMJ}
that published \cite{CF:happy} also published the
critique \cite{CCF:detect}. The cover of that issue, in fact, was devoted
to those two articles.
In contrast to {\it BMJ}'s decision,
the general-interest online newsmagazine {\it
Slate} published an article \cite{Johns:CF2}
on our critique the same month we submitted our paper.
An delightful coda is that a few months later, {\it BMJ} published an
editorial called ``Inadequate post-publication review of medical research"
\cite{inadeq}.

After these rejections by the {\it New Engl.\ J.\ Med.} and {\it BMJ}, we
approached three top journals who did not publish any of C\&F's
studies, {\it JAMA}, {\it Lancet}, and {\it
Proc.\ Natl.\ Acad.\ Sci.}. All were uninterested in our critique
since they do not publish
critiques of articles they did not originally publish. The section of
{\it J. Pers.\ Soc.\ Psychol.}\ that published \cite{CFC:alone} does not
publish critiques even of papers they have published, unless accompanied by
new data.

Following on this educational venture,
we submitted to a statistics journal that specializes in reviews, {\it
Stat.\ Sci.} Five months later they had 3 referee reports. The first
two recommended publication after revisions (e.g., ``an important critique"
and ``well worth publishing"), while the third, though agreeing with our
critiques, said that C\&F's work was insufficiently important
to warrant publication of a critique in {\it Stat.\ Sci.} Two months
after getting these reports, the editor made his decision: rejection,
allowing for resubmission if we made the tone more neutral and changed the
focus, perhaps to ``editorial decision making standards in medical
journals", as suggested by the third referee.

Methodological journals abound, but their cautions and recommendations are
largely ignored \cite{Blalock}.
Indeed, ``in a process well documented by Blalock and Duncan, positivist
sociology, like so many other professions, has tended to become immune to
the recognition of flaws in its work" \cite{twilight}.
Given the above considerations,
it may help to have a journal specifically devoted to
critiques.\footnote{We thank Elchanan Mossel for this idea.} 
This would not only allow others to know more about which studies are
trustworthy, but could also have the salutary effect of encouraging
researchers to pay extra attention to their methods lest they be publicly
critiqued.


\bsection{Conclusions}
\label{s.conclude}

We begin by summarizing the major problems
with C\&F's studies:
\begin{enumerate}
\item The data are not available to others.
\item The unavailable data are sparse for friendships.
\item The models used to analyze the sparse data contradict
the data and the conclusions.
\item The method used to estimate the dubious 
models does not apply.
\item The statistical significance tests from the questionable
estimates do not show the proposed differences.
\item The wrongly proposed differences do not distinguish among
homophily, environment, and induction.
\item Associations at a distance are better explained by
homophily than by induction.
\end{enumerate}

%

How did these errors arise and pass inspection? 
We believe that one major reason is that,
as many before us have said,
statistical assumptions
are routinely made when they are unlikely to hold.
The motivation for making assumptions is the hope of
overcoming the limitations of observational data, especially for causal
inference.
In any particular case, some of those limitations are known, while others
are unknown.
Yet viewing observational data through the lens of
statistical modeling produces new biases, generally unknown and
mostly unacknowledged, lurking in mathematical thickets.
Unfortunately, controlling for selection
effects and other confounders
is extraordinarily difficult in observational studies
\cite{contradicted}; this is the
main reason that observational studies are regarded with
skepticism.
Indeed, as demonstrated with the well-known studies concerning
hormone-replacement therapy, it was impossible to control
the observational studies to get the
same effects as the experiments \cite{PF:how,FP:HRT}.
Observational studies often lead to publications whose causal conclusions
contradict one
another or are contradicted by experiments
\cite{contradicted,triumph,Taubes}; this is a natural
consequence of poor methodology. 

Some investigators have found much better data, even if not perfect, to
assess causal effects in social networks.
For example, \cite{Sacerdote,CHW}
look at random assignments of college freshmen. This is promising, although
both these cited studies make assumptions about what their data look like
without presenting that data, nor telling us why the data fit their models.
It is not uncommon to subject experimental data to statistical modeling,
but this, too, likely leads to biases \cite{DAF:adj}. 
Of course, 
small-scale experiments could be initiated to see what the effects of
intervention actually are. 
Since the collection of good data is usually very hard and expensive, most
papers substitute for it by statistical modeling \cite{Blalock}.
``But  progress  is unlikely," wrote Summers \cite{Summers},
``as  long  as  [we]  require  the
armor  of  a  stochastic pseudo-world before doing battle with evidence
from the real one."

We can learn \cite{SM:TaP,survival}
from others' experiences of modeling observational data,
but there is also an {\it a priori} reason to distrust modeling
in the absence of the ability to confirm or deny the results:
Rarely can one know whether
the needed assumptions are correct---otherwise, they
wouldn't be assumptions \cite{oasis}.
Yet they are crucial to analysis when modeling.
In order to bring these issues to light in published research, analysts
who model should, at a minimum, 
state their models fully and explicitly, complete with equations
and assumptions. Similarly, analysts should clearly report their estimation
methods and the assumptions behind them.
Such clarity will not only aid readers and reviewers, it may also alert
authors to mistakes in reasoning before they are committed to print.
It is also wise to bear in mind that  
technical fixes (such as adding a lagged obesity term to a logistic
regression model to account for homophily)
work only for technical problems, not for fundamental
issues \cite{SM:TaP}.
Other recommendations for better statistical practice can be found in
\cite{survival,Altman} (among hundreds of other articles).

Medical journals often publish articles and
editorials measuring and bemoaning the quality of 
evidence in medicine \cite{Glantz,wisdom,Altman}.
The situation is so bad that
a recent study \cite{mostfalse} of the medical literature
was titled ``Why Most
Published Research Findings Are False"\footnote{Although the title may
be correct, the article did not provide sufficient empirical
evidence to establish
it.}; the author later was appointed to
the Stanford University School of Medicine.
These concerns have spread to the popular media: 
see \cite{oddsare,DHFreedman,Begley}.
Of course, the nature of almost all medical evidence is statistical.



Medicine has many comrades who share its concern over the misuse of statistics.
Since Keynes \cite{Keynes1,Keynes2},
there have been 
some social scientists who have decried tendencies in their fields towards
statistical idolatry; Keynes used the phrases ``black magic" and
``statistical alchemy".
See \cite[Chap.\ 10]{SM:TaP} for a review of some of the vast
critical literature and, e.g., \cite{Gigerenzer,twilight} and the
references there.
Among the concerned was
Otis Dudley Duncan, one of the most important quantitative
sociologists of the last century. More than 25 years ago, he
wrote \cite[p.~226]{ODD}:

\technical

\noindent
Coupled with downright incompetence in statistics, paradoxically, we often
find the syndrome that I have come to call statisticism: the notion that
computing is synonymous with doing research, the naive faith that
statistics is a complete or sufficient basis for scientific methodology,
the superstition that statistical formulas exist for evaluating such things
as the relative merits of different substantive theories or the
``importance" of the causes of a ``dependent variable"; and the delusion
that decomposing the covariations of some arbitrary and haphazardly
assembled collection of variables can somehow justify not only a ``causal
model" but also, praise the mark, a ``measurement model."  There would be no
point in deploring such caricatures of the scientific enterprise if there
were a clearly identifiable sector of social science research wherein such
fallacies were clearly recognized and emphatically out of bounds.

\endtechnical

\noindent
Duncan hoped that criticism of such ``abuses \dots\ might lead to
something like the famed Flexner report of 1910 that put the spotlight on
the miserable state of medical education at that time." \cite[p.\ 227]{ODD}
Indeed, an obvious ``cure" for poor statistical practice is to improve
statistics education.
While there is widespread agreement on the need for statistical literacy
among the populace at large, efforts to improve the statistical competence
of those who become practitioners receive less attention.

We see the problems with existing statistics education as follows.
Although most statistics courses mention the importance of the assumptions
behind the techniques they present, few devote much time to this topic.
Such lack of attention is
especially prevalent in more advanced courses taught in a variety of
disciplines\footnote{Those who feel that
most textbooks do pay serious attention to assumptions are urged to compare
those books to \cite{FPP,SM:TaP} and, especially, to compare their
exercises. Especially useful are exercises that present possible mistakes
in the published literature, while asking students whether the statistical
conclusions were justified.},
yet the assumptions behind more advanced
techniques are considerably more subtle than those in elementary courses. 
Most students, who are generally practically minded,
learn not to question whether the assumptions hold in practical
situations---or, at least, students do not learn to question the
assumptions. Many such students
later become practitioners and, often, educators themselves: more
statistics is taught outside statistics departments than within.
In the face of academic pressure to publish papers, assumptions become
inconvenient and further marginalized, even though all assent to their
importance.
Thus, Blalock \cite{Blalock} wrote of his profession, sociology:

\technical

\noindent
[O]ne finds a large number of journal articles that briefly discuss the
measurement of selected variables, that also admit to the possibility of
errors, but that then effectively announce to the reader that the
subsequent empirical analysis and related interpretations will proceed as
though there were absolutely no measurement errors whatsoever! 

This is but one illustration of the more general point that
methodological ideas are adopted when it is relatively easy and costless
to do so, but that they are resisted or totally ignored when it is to the
investigator's vested interest to do so.

\endtechnical

\noindent
It is to counteract these natural tendencies that we urge much greater
attention to questioning assumptions.

Flawed statistical models are not limited to medicine and
academia. For example, the current
economic afflictions are partly due to flawed models: see
\cite[Section II.D]{Stig}, \cite[pp.~16, 28, 44, 149]{FCIC} and
\cite[pp.~288ff]{senate}.
An examination of statistics education, as Duncan suggested, is overdue.
Educators need not wait for any report, however, before we ourselves teach
critical thinking \cite{FPP,SM:TaP}.

\medbreak
\noindent {\bf Acknowledgements.}\enspace I am grateful to Abie Flaxman,
Jason Fletcher, Elizabeth Housworth, Janet Macher, Roger Purves, Philip
Stark, and Duncan Watts for helpful conversations, suggestions, and
remarks. I also thank the referees for useful suggestions and references.

\bigbreak

\appendix

\section{Directionality Table}
\label{a.dtable}

The overlapping confidence intervals for directional coefficient 
estimates were shown
in \rref f.CIslabs/. The actual numbers are given here in Table 1. They are
reported both as probability estimates with CIs and as coefficient estimates
with SEs, for the following reason.
Logistic regression models transform numbers on the right-hand side into
probabilities on the left-hand side. However, one must choose values for
every covariate in order to get a probability.
Even when one varies a right-hand side coefficient in order to see how the
uncertainty in its estimate transforms into an uncertainty in
probability, one must choose values for all the covariates because of
the non-linear nature of the transformation.
Since this transformation depends on the values chosen for the
covariates, there is in principle one probability and one CI for each
FP. What C\&F report instead are probabilities and CIs when the covariates
are assigned their mean values over the population. This doesn't represent
anyone (e.g., the gender variable is 1/2, while for a person, it is either
0 or 1).  Thus, such
probabilities and CIs are only a vague kind of average of the individual
probabilities and CIs. This is a well-known
difficulty with logistic regression
models.

\begin{figure}[!htp]
$$
\vcenter{\offinterlineskip
\eightpoint
\def\bigstrut{\vrule height12pt depth5pt width0pt}
\hrule
\halign{\vrule\bigstrut\quad\hfil #\hfil\quad&
  \vrule\quad\hfil #\hfil\quad&
  \quad\hfil #\hfil\quad&
  \quad\hfil #\hfil\quad&
  \vrule#
  \cr
Source&\mut&\fplp&\lpfp&\cr
\noalign{\hrule}
\rref b.CF:obes/, p.~376&
171\% [59\%, 326\%]&
57\% [6\%, 123\%]&
13\% $[-28\%, 68\%]$&
\cr
\rref b.CF:obes/, suppl.~p.~3&
$1.19 \ (0.33)$&
$0.52 \ (0.23)$&
$0.11 \ (0.28)$&
\cr
\rref b.CF:peer/, p.~1401&
&
$0.033 \ (0.014)$&
$0.002 \ (0.014)$&
\cr
\rref b.CF:smoke/, pp.~2254, 2256&
43\% [1\%, 69\%]&
36\% [12\%, 55\%]&
15\% $[-35\%, 50\%]$&
\cr
\rref b.CF:smoke/, suppl.~p.18&
$0.66 \ (0.33)$&
$0.51 \ (0.19)$&
$0.21 \ (0.27)$&
\cr
\rref b.CF:happy/, p.~6&
63\% [12\%, 148\%]&
25\% [1\%, 57\%]&
12\% $[-13\%, 47\%]$&
\cr
\rref b.CF:happy/, suppl.~p.~9&
$2.07 \ (0.79)$&
$0.70 \ (0.34)$&
$0.32 \ (0.41)$&
\cr
\rref b.CFC:alone/, pp.~983--984&
$0.41 \ (0.13)$&
$0.29 \ (0.11)$&
$0.35 \ (0.30)$&
\cr
}
\hrule}
$$
\figcaption{5.7}
{{\bf Table 1.} Directional differences for friendship ties.
Key: 
\mut\ means mutual friendship;
\fplp\ means FP named LP; \lpfp\ means LP named FP;
FP = ego; LP = alter.
[Reported 95\% CIs] and (reported SEs).}
\medskip
\end{figure}

\bsection{Further Lack of Statistical Significance}
\label{s.further}

\rref s.direct/ showed that
C\&F's directional analysis was flawed by lack of
statistical significance (among other flaws). This same flaw occurs in
other comparisons C\&F make.  For example, \cite{CF:happy} states that
``Coresident spouses who become happy increase the probability their spouse
is happy by 8\% (0.2\% to 16\%), while non-coresident spouses have no
significant effect." That is, C\&F say that coresident spouses have an
effect, while non-coresident spouses do not. The mistake is that this is
based on the second covariate (non-coresident spouses) having a coefficient
that is statistically non-significant: the coefficient translates to a
probability of 2\% with a CI so large,
$[-18\%, 31\%]$, that it engulfs the CI for the first covariate.
Thus, the {\em difference} between the
two coefficients cannot be said to be statistically significant.  Again,
C\&F's methods do not permit a comparison between the importance of these
two covariates. 
Similar examples are listed in Table 2.

\begin{figure}[!htp]
$$
\vcenter{\offinterlineskip
\eightpoint
\def\bigstrut{\vrule height12pt depth5pt width0pt}
\hrule
\halign{\vrule\bigstrut\quad\hfil#\hfil\quad&
  \vrule \quad\hfil #\hfil\quad&
  \quad\hfil #\hfil\quad&
  \vrule #
  \cr
Source&Covariate 1&Covariate 2&\cr
\noalign{\hrule}
\rref b.CF:obes/, p.~376&
same sex 71\% [13\%, 145\%]&
opposite sex $-9\%$ $[-62\%, 117\%]$&
\cr
\rref b.CF:obes/, p.~376 &
M same sex 100\% [26\%, 197\%]&
F same sex 38\% $[-39\%, 161\%]$&
\cr
\rref b.CF:smoke/, p.~2254&
FP college 57\% [29\%, 75\%]&
LP no college 4\% $[-67\%, 43\%]$&
\cr
\rref b.CF:smoke/, p.~2254&
LP college 55\% [26\%, 74\%]&
LP no college 4\% $[-67\%, 43\%]$&
\cr
\rref b.CF:smoke/, p.~2254&
both college 61\% [28\%, 81\%]&
LP no college 4\% $[-67\%, 43\%]$&
\cr
\rref b.CF:smoke/, pp.~2255--2256, suppl.\ p.~31&
moderate smoking, various&
heavy smoking, various&
\cr
\rref b.CF:smoke/, suppl.\ p.~15&
late period $-70.89 \ (35.9)$&
early period $11.49 \ (13.3)$&
\cr
\rref b.CF:happy/, p.~6&
nearby friend 25\% [1\%, 57\%]&
distant friend $-3\%$ $[-15\%, 10\%]$&
\cr
\rref b.CF:happy/, pp.~6--7&
coresident spouse 8\% [0.2\%, 16\%]&
non-coresident spouse 2\% $[-18\%, 31\%]$&
\cr
\rref b.CF:happy/, pp.~6--7&
nearby sibling 14\% [1\%, 28\%]&
distant sibling 2\% $[-3\%, 8\%]$&
\cr
}
\hrule}
$$
\figcaption{4}
{{\bf Table 2.} Statistically insignificant comparisons.
Covariate 1 is statistically significant, while Covariate 2 is not.
[Reported 95\% CIs] and (reported SEs).}
\medskip
\end{figure}

Even when not making comparisons, C\&F
sometimes conclude that a number is 0 when their methods tell them
only that they cannot distinguish it statistically from 0. 
For example, \cite{CF:obes} states that ``Obesity in 
a sibling of the opposite sex did not affect the 
chance  that  the  other  sibling  would  become 
obese."
Other such examples are listed in Table 3.

\begin{figure}[!htp]
$$
\vcenter{\offinterlineskip
\eightpoint
\def\bigstrut{\vrule height12pt depth5pt width0pt}
\hrule
\halign{\vrule\bigstrut\quad\hfil#\hfil\quad&
  \vrule \quad\hfil #\hfil\quad&
  \vrule #
  \cr
Source&Covariate&\cr
\noalign{\hrule}
\rref b.CF:obes/, p.~376&
opposite sex sibling 27\% [3\%, 54\%]&
\cr
\rref b.CF:smoke/, suppl.\ p.~15&
early current centrality $2.20 \ (91.31)$&
\cr
\rref b.CF:smoke/, suppl.\ p.~15&
late current centrality $-138.00 \ (156.00)$&
\cr
\rref b.CF:happy/, p.~6, suppl.~p.~7&
additional unhappy alter $-0.06 \ (0.03)$&
\cr
\rref b.CF:happy/, p.~7, suppl.~p.~10&
coworkers $-0.29 \ (0.16)$&
\cr
}
\hrule}
$$
\smallskip
\figcaption{5.5}
{{\bf Table 3.} Statistically insignificant conclusions.
Covariate coefficient is reported not statistically significant, but the
authors treat it as 0, even though the CI was not close to 0.
[Reported 95\% CI] and (reported SEs).}
\medskip
\end{figure}

\bibliography{CF}
\bibliographystyle{ieeetr}

\end{document}

%% file: russ2latex.tex
\usepackage[usenames]{color}
\usepackage{graphicx}
\usepackage[headings]{fullpage}
\usepackage[font=small,labelfont=bf,labelsep=period,width=4.5in]{caption}

\usepackage[colorlinks=true,citecolor=OliveGreen,linkcolor=BrickRed,urlcolor=BrickRed]{hyperref}

\def\thmenv#1#2#3{\begin{#1} \label{#1:#2} #3 \end{#1}}

\def\procl#1.#2 #3\endprocl{%
       \ifx#1t\thmenv{thm}{#2}{#3}\fi
       \ifx#1l\thmenv{lem}{#2}{#3}\fi
       \ifx#1p\thmenv{prop}{#2}{#3}\fi
       \ifx#1c\thmenv{cor}{#2}{#3}\fi
       \ifx#1d\thmenv{defn}{#2}{#3}\fi
       \ifx#1g\thmenv{conj}{#2}{#3}\fi
       \ifx#1q\thmenv{quest}{#2}{#3}\fi
       \ifx#1r\thmenv{rmk}{#2}{{\rm #3}}\fi
    }%

\def\rref#1.#2/{%
      \ifx #1sSection~\ref{s.#2}\fi
      \ifx #1tTheorem~\ref{t.#2}\fi  
      \ifx #1lLemma~\ref{l.#2}\fi 
      \ifx #1cCorollary~\ref{c.#2}\fi 
      \ifx #1pProposition~\ref{p.#2}\fi 
      \ifx #1dDefinition~\ref{d.#2}\fi
      \ifx #1gConjecture~\ref{g.#2}\fi 
      \ifx #1qQuestion~\ref{q.#2}\fi 
      \ifx #1aAppendix~\ref{a.#2}\fi 
      \ifx #1fFigure~\ref{f.#2}\fi
      \ifx #1e(\ref{e.#2})\fi
      \ifx #1b\cite{#2}\fi
        }

\def\rlabel #1 #2{\begin{equation} \label{eq:#1} #2 \end{equation}}

\def\proofof #1.#2 {\medbreak\noindent
     {\it Proof of \rref #1.#2/.}\enspace}

\def\eightpoint{\footnotesize}
\def\technical{\medbreak\begingroup\small\narrower\narrower}
\def\endtechnical{\smallbreak\endgroup\noindent}
\def\figcaption#1#2{\parbox{#1 truein}{\eightpoint {#2}}}

\def\bsection{\bigbreak\section}
